\documentclass[12pt]{extarticle}
\usepackage[utf8]{inputenc}
\usepackage[T1]{fontenc}
\usepackage[a4paper, left = 2.5cm,right = 2.5cm,top = 2.5cm,bottom = 3cm]{geometry}
\usepackage{amsmath,amssymb}
\usepackage{cite}
\usepackage{hyperref}
\usepackage{newtxtext,newtxmath}
\begin{document}
	\title{Explicit isometric embeddings of pseudo-Riemannian manifolds: ideas and applications}
\author{A. A. Sheykin\thanks{e-mail:anton.shejkin@gmail.com}, \ M. V. Markov\thanks{e-mail:vkrms321@gmail.com}, \ Ya. A. Fedulov\thanks{e-mail:yar\_{}00@inbox.ru} \ and S. A. Paston\thanks{e-mail:pastonsergey@gmail.com}\\
	{\it Saint Petersburg State University, Saint Petersburg, Russia}}
\date{}
\maketitle

\begin{abstract}
We study the problem of construction of explicit isometric embeddings of (pseudo)-Riemannian manifolds. We discuss the method, which is based on the idea that the exterior symmetry of the embedded surface and the interior symmetry of its metric must be the same. In case of high enough symmetry of the metric such method allows transforming the expression for induced metric, which is the one to be solved in order to construct an embedding, into a system of ODEs. It turns out that this method can be generalized to allow the surface to have lower symmetry as long as the above simplification occurs. This generalization can be used in the construction of embeddings for metrics, whose symmetry group is hard to analyze, and the construction of the isometrically deformed (bent) surface. We give some examples of the application of this method. In particular, we construct the embedding of spatially-flat Friedmann model and isometric bendings of a sphere, 3-sphere, and squashed AdS universe, which is related to the Godel universe. 
\end{abstract}

\section{Introduction}
An isometric embedding of an $n$-dimensional (pseudo)-Riemannian spacetime is a procedure of finding a surface in $N$-dimensional ambient spacetime (which is often assumed to be flat) the metric on which coincides with metric of the original spacetime:
\begin{align}\label{metric}
g_{\mu\nu}(x)=\partial_\mu y^a (x) \partial_\nu y^b (x) \eta_{ab},
\end{align}
where $x^\mu$ are coordinates on the manifold, $y^a(x^\mu)$ is the so-called embedding function and $\eta_{ab}$ is a metric of ambient space, $\mu=0\ldots n-1$, $a=0\ldots N-1$. Janet-Cartan-Friedman \cite{fridman61} theorem states that for local embedding, it is sufficient to have $N={n(n+1)}/{2}$, although symmetric spacetimes can be embedded in an ambient spacetime of lower dimension. The procedure of isometric embedding can serve many purposes, including the classification of solutions of Einstein equations \cite{schmutzer}, investigation of their global structure (in particular, black holes \cite{frons}), calculation of Hawking temperature of spacetimes with a horizon \cite{deserlev98,deserlev99,statja34,statja36} and even the modification of gravity \cite{regge,deser,pavsic85let,davkar,statja25,faddeev,statja51}. In many of these studies, it is often necessary to have an explicit embedding of the considered spacetime. This can be done by solving the system \eqref{metric} w.r.t. $y^a$. Unfortunately, the system \eqref{metric} is a nonlinear system of PDEs, so it is difficult to solve it straightforwardly, and the regular methods of solving it in the generic case are scarce.

There is, however, a method of construction of explicit isometric embeddings for Riemannian manifolds, which possesses some symmetry \cite{statja27}. The idea is to construct a surface, which transforms into itself under the action of some subgroup of the Poincare group of ambient spacetime, which is isomorphic to the symmetry group of the metric. This method was then successfully applied to many spacetimes, mainly black hole ones \cite{statja27,statja29,statja30,statja40,statja56,statja57}. However, some problems cannot be solved by directly applying this method: the construction of an isometrically deformed surface that ought to have different symmetry than the metric. Therefore we propose a modification of it:
if one is unable to (or does not want to) construct an embedding function with the same symmetry as the metric (in the above sense), one might try to lower the symmetry of the embedding function. The disadvantage of this modification is the fact that the system \eqref{metric} becomes more difficult to solve. Nevertheless, as long as one can separate variables in it, the problem remains solvable in principle.

In this paper, we illustrate this idea with some simple examples. Section 2 is devoted to the problem of isometric bending of 2D and 3D spheres. In section 3, we obtain the embedding for the spatially flat Friedmann model (which is notoriously difficult to construct compared to two other models). In section 4, we construct a new embedding for the deformed Godel universe, and in section 5, we discuss some properties of embeddings obtained using this method.

\section{Sphere}
Let us consider an $SO(3)$-symmetric metric (2-sphere):
\begin{align}
ds^2 =  d\theta^2 + \sin^2 \theta d\phi^2.
\end{align}
There is only one embedding in 3D which has the same $SO(3)$ symmetry:
\begin{align}
 y^1  =  \sin\theta \cos \phi, \quad y^2  =  \sin\theta \sin \phi, \quad y^3  =  \cos\theta.
\end{align}
However, it is not the most general 2D surface with $SO(3)$-symmetric metric in 3D ambient space: one can bend a 2-sphere isometrically to obtain, e.g., a spindle-like surface. To find the family of surfaces which correspond to an isometric bending of the sphere let us lower the symmetry of embedding function from $SO(3)$ to $SO(2)$ and consider the following ansatz\cite{rashevsky}:
\begin{align}\label{2D}
y^1 = \Theta(\theta) \cos \Phi(\phi), \quad y^2 = \Theta(\theta) \sin \Phi(\phi), \quad y^3 = f(\theta).
\end{align}
Solving the system \eqref{metric} w.r.t. $y$, one obtains 
\begin{align}\Phi(\phi)=a\phi, \quad \Theta(\theta)=a^{-1}\sin\theta, \quad f(\theta)=\int d\theta \sqrt{1-\cos^2(\theta)/a^2}.\end{align} 
If $\phi\in[0,2\pi/a]$, it defines a family of spindle-like surfaces. In the codimension 1 it is the only $SO(2)$-symmetric solution for 2-sphere.

In higher dimensions, however, the situation is different: consider an $SO(4)$-symmetric metric 
\begin{align}
ds^2 = d\chi^2+ \sin^2 \chi (d\theta^2 + \sin^2 \theta d\phi^2).
\end{align}
This symmetry can be broken at least in two ways: to $SO(3)$ and to $SO(2)\otimes SO(2)$ (one could lower the symmetry even more, e.g. to $SO(2)$, but in this case the system \eqref{metric} will remain a system of PDEs, so it cannot be solved straightforwardly, and we will not consider this case). The ansatz for an embedding function corresponding to the former case
\begin{align}
y^1 = X(\chi) \sin\Theta(\theta) \cos \Phi(\phi), \ y^2 = X(\chi) \sin\Theta(\theta) \sin\Phi(\phi), \ y^3 = X(\chi) \cos \Theta(\theta), \ y^4 = f(\chi)
\end{align}
does not give anything but ordinary 3-sphere embedding: \begin{align}X=\sin{\chi}, \ f=\cos{\chi}, \ \Theta=\theta, \ \Phi=\phi.\end{align}
 It is worth noting that the presence of 2 or more components which depend on $\chi$ only does not change the picture. The ansatz for $SO(2)\otimes SO(2)$ case if the metric is written in Hopf coordinates \cite{10.1177/0278364909352700} has the following form:
\begin{align}\label{3D}
y^1 = X(\chi)  \cos \Phi(\phi), \  y^2 = X(\chi)  \sin\Phi( \phi), \  y^3 = h(\chi) \sin\Theta(\theta),    y^4 = h(\chi)\cos\Theta(\theta),   y^5 = f(\chi)
\end{align}
gives a 2-parametric family of surfaces:
\begin{align}X=a^{-1}\sin\chi, \ \ h=b^{-1}\cos{\chi}, \Phi=a\phi, \ \ \Theta=b\theta, \ \ f=\int d\theta \sqrt{1-\frac{1}{b^2}-\cos^2\theta\left(\frac{1}{a^2}-\frac{1}{b^2}\right)},\end{align} 
where $a^2\geq 1, b^2\geq 1$ for $f$ to be real-valued. Note that when $a=b=1$ the component $y^5$ is trivial and the rest of the ansatz \eqref{3D} components is a parametrization of Hopf coordinates. 

We must stress that while the problem of isometric bending of surfaces with constant curvature was widely discussed in the literature (both classic \cite{pogorelov} and modern \cite{ivanova3} one), in this paper we are interested in finding explicit expressions for embedding functions, so the discussion of existence of such embeddings is not our purpose here.   
\section{Spatially-flat Friedmann universe} 
Another metric to which the above method can be applied is the spatially-flat Friedmann universe: 
\begin{align}\label{fr}
ds^2 = dt^2 - a^2(t)(dr^2 + r^2 (d\theta^2 + \sin^2 \theta d\phi^2)).
\end{align}
It has $SO(3)\rtimes \mathbb{R}^3$ symmetry, i.e. the symmetry of 3-plane, but, unfortunately, there is no embedding function corresponding to \eqref{fr} whose $t=\text{const}$ sections are 3-planes. Therefore let us broke the $\mathbb{R}^3$ symmetry and consider an $SO(3)$-symmetric ansatz (signature is $(+----)$):
\begin{align}\label{f}
y^{0,1}=\frac{1}{2} (y^+(r,t) \pm y^-(r,t)), \ y^2  = a(t) r \cos\theta, \ y^3  = a(t) r \sin\theta \cos \phi, \ y^4  = a(t) r \sin\theta \sin \phi.	
\end{align}
In this ansatz the system \eqref{metric} is reduced to
\begin{align}
\dot{y}^+ \dot{y}^- - r^2 \dot{a}^2(t) = 1, \quad
{y^{+}}' {y^{-}}' =0, \quad \dot{y}^+{y^{-}}' + \dot{y}^- {y^{+}}' =2 r a(t) \dot{a}(t).
\end{align}
Since there are no mixed derivatives in this system, it can be immediately integrated to obtain
\begin{align}
y^{0,1}  = \frac{1}{2} \left(r^2 a(t) + \int \limits_0^t  \frac{dt}{\dot{a}(t)} \pm a(t)\right),
\end{align}
which gives the well-known Robertson embedding \cite{robertson1933}. It is also worth noting that it can be obtained as a limit $R\to \infty$ of closed or open universe embedding as well as through a classification of the representations of $SO(3)\rtimes \mathbb{R}^3$\cite{statja29}. 

\section{Godel universe}
The final example of application of modified method will be deformed Godel universe \cite{rooman} (hereafter we use dimensionless units in which  Godel metric parameter $a$ is equal to $1/2$, so $2a=1$):
\begin{align}\label{4}
{ds^2}=d{t}^2+2\mu\sinh^2 {\chi} dt d\phi-d{\chi}^2- (\sinh^2 {\chi} + (1-\mu^2)\sinh^4 {\chi}) d\phi^2-dz^2
\end{align}
($\mu^2=2$ corresponds to the Godel universe, whereas $\mu^2=1$ to the AdS one). The coordinates ${t}, {\chi}, {z} \in (-\infty,+\infty)$. The period of coordinate $\phi$ has to be $2\pi$ to avoid conical singularities. This fact is not obvious if the metric \eqref{4} is written in the coordinates $t,\chi,\phi,z$, but in other coordinates it can be shown \cite{rooman}. The symmetry of this metric in the case $\mu^2=2$ is $SO(2,1)\otimes SO(2)\otimes \mathbb{R}$ (the former is related to the coordinates $r,\phi$ and the latter to $t$)  \cite{rooman}. Note that at the level of algebra of Killing vectors it is impossible to draw a distinction between $SO(2)$ and $SO(1,1)$. Therefore the fact that Killing vectors of this metric  $\partial_t$ corresponds to $SO(2)$ can not be seen at the level of algebra. In principle, taking in mind the global structure of this metric, $SO(1,1)$, which is non-compact, could be the true $t$-related symmetry group. However, the famous existence of closed timelike curves in Godel universe favors $SO(2)$ \cite{rooman}.

Let us broke the $SO(2,1)$ subgroup of symmetry group to $SO(2)$ w.r.t. $\phi$ and consider the following ansatz (signature is $(1,\xi,-1,-1,\varepsilon,\varepsilon,-\varepsilon,-1)$, $\varepsilon=\text{sign}(\mu^2-1)$): 
\begin{align}\label{emb}
\begin{split}
 y^0=\sqrt{\xi}&  \frac{A(\chi)}{\alpha}\sin\left({\sqrt{\xi}}{\alpha}t\right), \  y^2 =  B(\chi) \sin (m\phi-\beta t), \ y^4 =   \frac{C(\chi)}{n} \sin(n\phi), \  y^6 = f(\chi),\\
 y^1=\xi& \frac{A(\chi)}{\alpha}\cos\left({\sqrt{\xi}}{\alpha}t\right),  y^3 =  B(\chi) \cos (m\phi-\beta t), \ y^5 =   \frac{C(\chi)}{n} \cos(n\phi), \  y^7 = z.
 \end{split}
\end{align}
This ansatz \textit{a priori} could give a local embedding of the metric \eqref{4}, but there are special cases when it describes a global one (see below). In the construction of embedding the $t$-related $SO(2)$ symmetry of metric can be represented either as $SO(2)$ symmetry of embedding function, so  $\xi=1$; or as $SO(1,1)$ one (this gives only local embedding), so  $\xi=-1$. In case when $\xi=-1$ the embedding function remains real-valued, but trigonometric functions in $y^{0,1}$ are transformed into hyperbolic ones. Note that other variants of representation of Abelian group by subgroups of Poincare group of ambient spacetime are also possible, but they require more dimensions \cite{statja27}. 

When the ansatz \eqref{emb} is used, the solution of
 \eqref{metric} is	
  \begin{align}\begin{split}  & A(\chi)=\sqrt{1+ \frac{\mu \beta}{m}  \sinh^2 \chi}, \quad B(\chi)=\sqrt{\frac{\mu}{m\beta}}\sinh \chi, \\ & C(\chi)=\sqrt{\varepsilon h(\chi)} \sinh \chi, \qquad h(\chi)=\left(\frac{m\mu}{\beta}-1\right)+(\mu^2-1)\sinh^2 \chi,\\ & f(\chi) = \int d\chi  \sqrt{\varepsilon \left(1-\frac{\mu}{m\beta} \cosh^2 \chi + \frac{\xi}{\alpha^2}A'^2  +\frac{h}{n^2}\left(\cosh \chi + \left(\frac{h'}{2h}\right)\sinh \chi \right)^2 \right)}.\end{split}\end{align}
   It defines a family of surfaces which are parametrized by constants $m, n, \alpha$ and $\beta$. In case when $m=n=1$ the surface possesses $SO(2)$ symmetry w.r.t. $\phi$ (with the period $2\pi$, which corresponds to the global symmetry of metric, see above), and if $\xi=1$ and $\beta/\alpha\in \mathbb Q$, there is $SO(2)$ symmetry w.r.t. coordinate $t$. In this case the embedding constructed here has global $SO(2)\otimes SO(2)$ symmetry which is a part of $SO(2,1)\otimes SO(2)$ symmetry of metric \eqref{4}.  If we, in addition, suppose that $\alpha=\beta=1/\mu$, an interesting special case appears, for which \begin{align}\begin{split} &A(\chi)=\cosh{ \chi}, \qquad \qquad \quad B(\chi)=\mu \sinh \chi, \ \\ &C(\chi)= \frac{\sqrt{|\mu^2-1|}}{2} \sinh  2\chi,  \ \ f(\chi) = \frac{\sqrt{|\mu^2-1|}}{2}\cosh 2\chi ,\end{split} \end{align} so $y^{0,1,2,3}$ and $y^{4,5,6}$ blocks define two quadratic surfaces. This particular case was found in \cite{rooman}, and at those values of parameters the resulting embedding is global. Otherwise, conical singularities could be present. It is interesting to note that for Godel universe ($\mu^2=2$) the expressions also simplify when $m=n=1$, $\alpha=1/\sqrt{2}$, $\beta=\sqrt{2}$, but the symmetry of the corresponding surface is unclear.

\section{Discussion}
In this paper, we demonstrate how the method of constructing explicit isometric embeddings of Riemannian manifolds can be generalized to allow the studying of isometrically deformed surfaces. It can be done in the following steps:
\begin{itemize}
	\item Identify the symmetry group of a metric.
	\item Determine whether it is possible to convert the system of PDE \eqref{metric} into the system of ODE by choosing a symmetric surface as an ansatz.
	\item If so, lower the symmetry of metric to the smallest possible group which allows it.
	\item Solve the remaining ODEs and check the globality of the constructed embedding.  
\end{itemize}

 The presence of central Abelian subgroups in the symmetry groups of deformed surfaces with metric of 2D sphere \eqref{2D}, 3D sphere \eqref{3D} sphere and deformed Godel universe \eqref{emb} (as well as the absence of those in the symmetry group of embedding function \eqref{f} of the Friedmann universe) gives a hunch that there might be some connection between the dimension of central Abelian subgroup of the symmetry group of the surface and the number of bending parameters of the surface. It can be conjectured that the surface, which has a $p$-dimensional center in its symmetry group, allows $q$-parametric bending with $q\geq p$. This conjecture arises from the fact that in the realization of each of the Abelian subgroups in the center of the symmetry group through a translation and a rotation in the ambient space, arbitrary constant appears as multipliers of coordinate which corresponds to the action of this one-parametric subgroup. For example, $\alpha$ and $\beta$ are constants corresponding to the $t$-related $SO(2)$ subgroup, whereas $m$ and $n$ --- to the $\phi$-related one. The situation $q=p$ occurs in the case when each of the central subgroups is realized irreducibly. The investigation of the validity of this conjecture is beyond the scope of the present paper.

{\bf{Acknowledgments}}.{The work of A. S. and S. P. is supported by RFBR Grant No.~20-01-00081.
	The authors are grateful to	 D.~P.~Solovyev for the useful references and to A.~N.~Starodubtsev for valuable discussions.}

\end{document}